\documentclass[prb,showpacs,preprintnumbers,amsmath,amssymb]{revtex4}
\usepackage{graphicx}% Include figure files
\usepackage{dcolumn}% Align table columns on decimal point
\usepackage{bm}% bold math
\usepackage{subfigure}

\begin{document}
\title{5. Skyrmion and helical states in thin layers of magnets and liquid crystals \label{spherulites}}

\email{leon-off@mail.ru}

% %%% PACS numbers

%\keywords{surface-induced anisotropy, ferromagnetic nanowires, nanotubes, vortex states of the magnetization}%Use showkeys class option if keyword
                              %display desired
         
\maketitle

%%%%%%%%%%%%%%%%%%%%%%%%%%%%%%%%%%%%%%%%%%%%%%%%%%%%%%%%%%%%%%%%%%%%%%%%%%%%%%%

\section{Introduction}

As shown in chapter 4, specific chiral interactions in \textit{bulk} non - centrosymmetric helimagnets provide a unique mechanism for stabilization of the localized and modulated structures with fixed sense of the magnetization rotation in one (spirals) or in two (baby-skyrmion) dimensions. Twisted modulations of such a type arise as a result of the competition between the exchange stiffness and chiral interactions, whereas Zeeman energy in accordance with smaller energy contributions  %(in the thesis I consider the influence of uniaxial, cubic, and exchange anisotropies on Skyrmion states) 
rules the thermodynamical stability of the corresponding modulated phases. % collapse into the singularities for centrosymmetric systems. 
Currently however, the scope of material science interest has been shifted to the \textit{confined} artificial systems in which the structural inversion asymmetry of the surface gives rise to qualitatively new phenomena. The new aspects of the physics in reduced dimensions include Rashba effect \cite{Koroteev04,Hirahara06}, chiral Dzyaloshinskii-Moriya interactions \cite{PRL01,Bode07}, surface/interface induced anisotropy \cite{Bogdanov02r,Neel54}, and/or anchoring in liquid crystals. Understanding of these effects opens up the perspectives to create chiral architectures in nanosystems and put into practice control over them.

Surface-induced uniaxial anisotropy is a key factor that can influence chiral modulated states in magnetic nanosystems. Its interplay with volume energy contributions leads to the formation of specific axisymmetric distributions of the magnetization, spherulites, which exist as smooth static solitonic textures and are extended into the third direction in accordance with the modulating effect of the surfaces. As a starting point for investigations of this phenomenon I have calculated the equilibrium structures of skyrmions in a nanolayer with the induced uniaxial anisotropy for a case of "thick" layers  when the induced magnetic anisotropy is strongly localized to the layer surfaces (a rigorous criterion of "thick" layer can be found in Ref.\onlinecite{Bogdanov02r}. In this limiting case the induced anisotropy can be treated as a mere surface effect (according to N\'eel's theory of surface anisotropy \cite{Neel54}) and is included into the corresponding micromagnetic equations only through boundary conditions. In section \ref{isolatedspherulites}-\ref{spherulitesPD} I give a comprehensive analysis of the spherulitic states in confined nanolayers.

Dipolar stray fields appearing on the surface of confined layers are another important factor having a sizable effect on modulated states. Due to the strong magnetodipole interaction chiral helices and skyrmion lattices can be significantly deformed. In section \ref{SkMagnetostatic} I address this practically unexplored problem of the interplay between short-range energy contributions stabilizing modulated phases, skyrmions and helices, and the long-range magnetostatic forces. % formation of the equilibrium multidomain patterns under the competing influence of the Dzyaloshinskii-Moriya and magnetostatic interactions. 
%Interplay between these induced interactions with shortrange intrinsic couplings and long-range magnetostatic forces form complex spatial inhomogeneous magnetic structures including patterns with unbalanced chirality.

\section{Phenomenological model of modulated states in thin magnetic films \label{spherulitesPhen}}
Generally the surface-induced anisotropy must be considered as inhomogeneously distributed through the volume of the nanosystem. However, the case with the induced anisotropy constrained to the surface region represents the practically most important situation. When the penetration length $\lambda_s$  is much smaller then the characteristic sizes of the system, the surface-induced interactions influence the magnetization distributions only through the boundary conditions for the energy functional. In Refs. \onlinecite{Bogdanov02r,condmat}, for instance, the corresponding  estimates led to $\lambda=1.9$\AA\, for Co/Au films and $\lambda=26.4;31.9$ \AA \, for Ni/Cu multilayer systems.

Therefore, in many practically important cases  the magnetic energy of a nanosystem can be  simplified by reducing to a sum 
of the volume ($w_{\mathrm{v}}(\mathbf{r})$) and surface ($w_{\mathrm{s}}(\mathbf{r})$) constibutions: 
\begin{equation}
W=\int_{V}w_vd\mathbf{r}+\int_{S}w_{\mathrm{s}}d\mathbf{r}.
\end{equation}
Then the equilibrium distributions of the magnetization
$\mathbf{m}(\mathbf{r})$ are derived by solving the Euler equations for the volume functional $w_{\mathrm{v}} (\mathbf{m})$
with the boundary condintions imposed by surface energy $w_{\mathrm{s}}(\mathbf{r})$ (see e.g. \cite{PRL01,JMMM02}).

In the present  chapter I consider a magnetic nanolayer infinite in $x$- and $y$-directions and confined by parallel planar
surfaces at $z=\pm T/2$.
I consider a simplified model of a chiral ferromagnet  with energy densities:
\begin{align}
%w=\int_{V}(A\sum_{i,j}(\partial_i m_j)^2+Â \varepsilon_{ijk}m_i\delta_jm_k+w_{0})d\mathbf{r}+\int_{S}f_sd\mathbf{r}.
w_v=A(\mathrm{grad}\mathbf{m})^2-\mathbf{H}\cdot\mathbf{m}M -K_u(\mathbf{m}\cdot\mathbf{a})^2 + w_D ,
\label{functional2}
\end{align}
$w_s=-K_s (\mathbf{m}\cdot \mathbf{s})^2$, $A$ is
a stiffness constant, $\mathbf{H}$ is the applied magnetic
field, $\mathbf{m}$ = $\mathbf{M}/M$ is
the unity vector along the magnetization 
($M = |\mathbf{M}|$ ), 
$K_u > 0 $ and $K_s > 0$ are constants of volume and surface
induced uniaxial anisotropies correspondingly,
unity vectors $\mathbf{u}$ and $\mathbf{s}$
define these "easy" magnetization directions.
Chiral DM energy is composed from antisymmetric
invariants linear with respect to first 
spatial derivatives, so called Lifshitz
invariants 
$\mathcal{L}_{ij}^{(k)} = m_i \left(\partial m_j/\partial x_k \right)
-m_i \left(\partial m_j/\partial x_k \right).$
Particularly for cubic noncentrosymmetric magnets
\cite{Dz64,Bak80}
\begin{equation}
w_D=D\,(\mathcal{L}_{yx}^{(z)}+\mathcal{L}_{xz}^{(y)}+\mathcal{L}_{zy}^{(x)})
=D\,\mathbf{m}\cdot \mathrm{rot}\mathbf{m}
\label{rotor}
\end{equation}
For uniaxial noncentrosymmetric ferromagenets DM functionals
$w_D(\mathbf{m})$ are listed in \cite{JETP89}.

Importantly, such a model allows to describe also thin layers of chiral nematic liquid crystals sandwiched between two glass plates. %, can be also described by the model  (\ref{DMdens1}) with boundary conditions (\ref{functional2}).
The possible  distributions of the director $\bf{n(r)}$ in a bulk cholesteric liquid crystal placed in an electric field $\mathbf{E}=E \mathbf{e}$ are determined by the minimization of the Frank free energy \cite{Oswald00,books}:
\begin{align}
f_v=\frac{\mathit{K}_1}{2}(\rm{div}\,\mathbf{n})^2+\frac{\mathit{K}_2}{2}(\bf{n}\cdot\rm{rot}\,\mathbf{n}-q_0)^2
+\frac{\mathit{K}_3}{2}(\mathbf{n}\times\rm{rot}\,\mathbf{n})^2-\frac{\varepsilon_0}{2}(\mathbf{n}\cdot\mathbf{E})^2. 
\label{Franc}
\end{align}
Here, $K_i\, (i=1,2,3)$ and $q_0$ are elastic constants; $\varepsilon_a$ is the dielectric anisotropic constant
(we consider only materials with $\varepsilon_a>0$);  
%$f_0=K/2-\varepsilon_a E^2/2$ is a free energy density
%for a homogeneous state with  $\mathbf{n}||\mathbf{E}$; 
$p=2\pi/|q_0|$ determines the pitch of  a helical
structure in the ground state (at $E=0$).
For the one elastic constant approximation, $K_1=K_2=K_3=K$, the Frank energy density (\ref{Franc}) can be reduced to the form:
\begin{equation}
f_v=K(\mathrm{grad}\,\mathbf{n})^2+2Kq_0\mathbf{n}\cdot{\mathrm{rot}}\,\mathbf{n}-\varepsilon_0E^2 (\mathbf{n}\cdot\mathbf{e})^2.
\label{lc1}
\end{equation}
For $\mathbf{H} =0$ the energy density $w_v$  coincides functionally with $f_v (\mathbf{n})$ (Eq. \ref{lc1}). This allows to investigate skyrmionic states in chiral ferromagnets and liquid crystals within a common mathematical
framework.
The surface energy $w_s$ for liquid crystals (Eq. (\ref{functional2})) describes a homeotropic anchoring of glass surfaces with $K_s > 0$ \cite{books}. In the experiments with liquid crystals, such glass surfaces confine a thin layer of a liquid crystal and anchor the molecules perpendicularly to the surfaces.
Phenomenological model (Eq. (\ref{lc1})) allows to obtain continuous distributions of the director $\mathbf{n}$ for liquid crystals. %This model is alternative to the conventional disclination models as it is a feature of liquid crystals to form complex textures with point and line defects \cite{Oswald00}. % according to which   

\section{Isolated skyrmionic states \label{isolatedspherulites}}
To investigate the structure of isolated skyrmions in thin magnetic layers one has to introduce  cylindrical coordinates for the spatial variable $\mathbf{r}=(r,\phi,z)$, and the spherical coordinates for the order parameters $\mathbf{m}$ and $\mathbf{n}$:  $\mathbf{m}, \mathbf{n} =(\sin\theta \cos\psi, \sin\theta \sin\psi, \cos\theta)$. On the contrary to the isolated skyrmions in bulk systems which are characterized by the dependence $\theta=\theta(r)$ in the cross-section of a skyrmion filement and are homogeneously extended into the third direction ($z$), the localized skyrmions in confined media are solutions of the following boundary value problem with  $\theta = \theta (r,z)$:
%
%As it was already explained  in section \ref{ISBH}, we introduce  cylindrical coordinates for the spatial variable $\mathbf{r}=(r,\phi,z)$, and the spherical coordinates for the order parameters $\mathbf{m}$ and $\mathbf{n}$: 
%$\mathbf{m}, \mathbf{n} =(\sin\theta \cos\psi, \sin\theta \sin\psi, \cos\theta)$.
%
\begin{align}
&\theta_{ZZ}+\theta_{\rho \rho}
+\frac{\theta_{\rho}}{\rho} - \frac{\sin \theta \cos \theta} {\rho^2}
-\frac{\sin^2 \theta} {\rho}
-\gamma\sin 2\theta +h\sin \theta =0,\nonumber\\
&(\theta_{Z}+\beta\sin\theta\cos\theta)|_{Z=\pm t/2}=0,\nonumber\\
&\theta(0,Z)=\pi, \theta(\infty,Z)=0.
\label{EulerSp}
\end{align}
Skyrmionic strings in such systems are distorted into specific 3D textures in accordance with the anchoring effect of the surfaces, but the topology remains the same \cite{JETP00}.

In Eq. (\ref{EulerSp}) I introduced the new length scales for magnets:
\begin{equation}
\rho_0=\frac{A}{D}
\end{equation}
and liquid crystals
\begin{equation}
\rho_0 = \frac{1}{2q_0}.
\end{equation}
Then $\rho=r/\rho_0, Z=z/\rho_0$.
Four control parameters in Eq. (\ref{EulerSp}), $\gamma$, $h$, $\beta$, and $t$,  are expressed as combinations of material parameters of the model functional (\ref{functional2}) and (\ref{lc1}).
For magnetic layers:
\begin{equation}
\gamma=\frac{AK_u}{D^2},  \;
 h=\frac{HM_s A}{D^2}, \; 
 \beta=\frac{K_s}{D}, \;
 t = \frac{TD}{A},
\end{equation}
For liquid crystals: 
\begin{equation}
\gamma=\frac{\varepsilon\varepsilon_0E^2}{K q_0^2},\, h=0, \; \beta= \frac{K_s}{2K q_0}, \; t = 2Tq_0.
\end{equation}%
Invariance of  Eq. (\ref{EulerSp}) under a scaling transformation:
\begin{equation}
t\rightarrow kt, \
\gamma\rightarrow \gamma k^2,\
 h\rightarrow h k^2, \
 \beta\rightarrow \beta k.
\label{scale}
\end{equation}
allows to investigate solutions of Eq. (\ref{EulerSp})
for a fixed reduced layer thickness $t$ ($t$=1 in the present chapter).
For the liquid crystals it is convenient to redefine
anisotropy parameter $\gamma$ and to write it in the units of $\varkappa$:
\begin{equation}
\gamma=\frac{\pi^2}{16\varkappa^2},\,\varkappa=\frac{E_0}{E},\,E_0=\pi q_0\sqrt{\frac{K}{4\varepsilon_0}}.
\end{equation}
The critical field $E_0$ \cite{JETP00} marks the crossover between localized and modulated skyrmion states
for bulk chiral helimagnets.
\begin{figure}
\centering
\includegraphics[width=16cm]{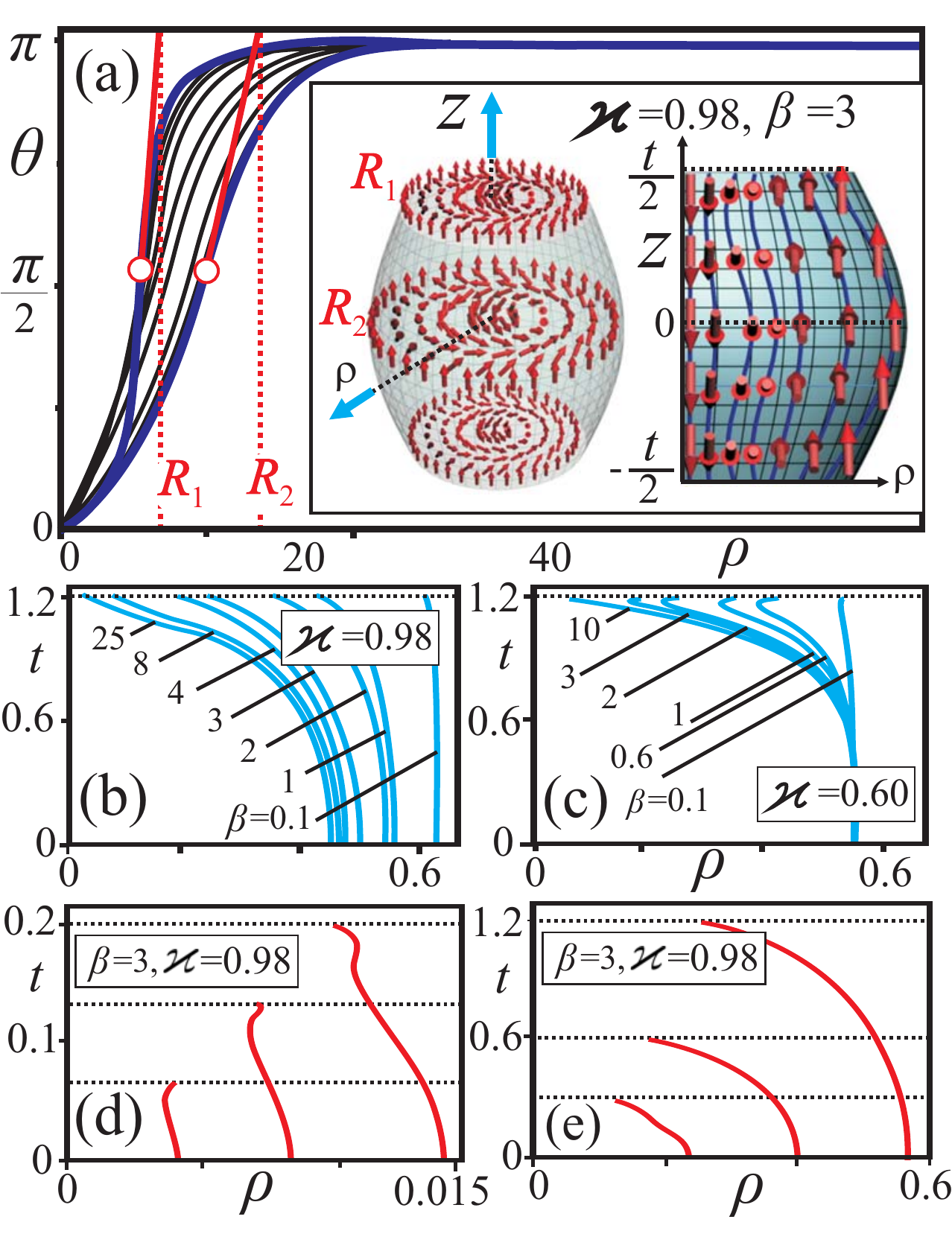}
\caption{
(a) Equilibrium solutions of isolated spherulitic domains,
shown as a set of profiles $\theta(\rho)$ for fixed coordinate $Z$ through the layer. Bold blue lines mark the profiles at the surface and in the center of the layer. 
Inset schematically displays the distribution
of the order parameter vector field with $\psi=\varphi-\pi/2$. 
The effect of anchoring is imaged
by its influence on the boundary surface of the spherulite
as shown in (a) by red lines for each angular profile:
(b), (c) represent shapes of the localized spherulites
for different values of surface anchoring $\beta$;
(d), (e) - for different film thicknesses $t$ .
\label{spherulitic}
}

\end{figure}

Boundary value problem (\ref{functional2}) has been
solved by a standard finite-difference method with
discretization on rectangular grids with adjustable grid
spacings.
As initial guess for the iterrative procedure according to Seidel method with Chebishev
acceleration \cite{Press07} I used the known solutions
of Eq. (\ref{functional2}) for bulk chiral systems \cite{JMMM94}, i.e. I started from the solutions with $\beta=0$.

Solutions for isolated skyrmions may be represented as a set of profiles $\theta(\rho)$ for fixed coordinates $Z$.
The profiles are strongly modified by the anchoring $\beta$ from bell-like type in the center of the layer to arrow-like at the surface (blue lines in Fig. \ref{spherulitic} (a)).
The thickness-dependent radii $R(Z)$ for each angular profile $\theta(\rho)$ are defined as shown by red tangent lines in Fig. \ref{spherulitic} (a), i.e. according to the Lilley definition. The function $R(Z)$ reproduces
the convex shape of solutions with the largest value in the center of the layer ($Z=0$) and the smallest value corresponding to
the layer surfaces ($Z=\pm t/2$) (Fig. \ref{spherulitic} (a), inset, and (b)-(e)).
With increasing anchoring parameter $\beta$ angular profiles $\theta(\rho)$ near the surfaces become strongly localized
(Fig. \ref{spherulitic} (b)) with the characteristic sizes comparable to the molecular length; in this case the elastic approach of Eq. (\ref{EulerSp}) is inapplicable, and the isolated skyrmions collapse into the homogeneous state.

The influence of the surface anchoring may lead not only to the compression of ideally cylindrical filement into convex-shaped spherulite, but also to the  specific skyrmionic states with "necks" when
the minimal value of $R(Z)$ is reached not at the surface of the layer, but at some coordinate $Z$
 in the volume (Fig. \ref{spherulitic} (c)). Such a peculiar shape of an isolated spherulite is a result of complex interplay between surface and volume energy contributions. It cannot be an artefact of the proposed method to define the function $R(Z)$ since the lines with constant angle $\theta$ show the same behaviour near the surface (see inset in Fig. \ref{spherulitic} (a)). % such a peculiar   shape of a spherulite may be related to the method of 

The decrease of the film thickness $t$ results in the same effects as increase of the anchoring parameter $\beta$ (Fig. \ref{spherulitic} (d), (e)) as it can be concluded from the equation (\ref{scale}) of the  scale transformation. % of Eq. (\ref{scale}) %of the Euler equation (\ref{Euler}) 
%the decrease of the film thickness $t$ results in the 
%same effects as increase of the anchoring parameter $\beta$ (Fig. \ref{spherulitic} (d), (e)). 
 %
The set of profiles $R(Z)$ for the fixed value of anchoring parameter $\beta=3$ and variable thicknesses $t$ of the layer displays the change of the spherulite shape from that with "necks" for thin layers to the convex shape for thick films (Fig. \ref{spherulitic} (d), (e)).

Still for the wide range of control parameters all angular profiles $\theta(\rho)$ are of arrow-like type squeezed along the sample thickness by anchoring. Such a type of angular profiles permits the  use of the ansatz \cite{JETP00}:
\begin{equation}
\theta(\rho,Z)=\pi\frac{\rho}{R(Z)}.
\label{ansatzR}
\end{equation}
The function $R(Z)$ describes the compression of angular profile $\theta(\rho)$ for $R > 1$ and its expansion with $R < 1$.
%
% Here, $R(z)$ is a positive number and characterizes the size of the profile core \cite{JETP00}.
%
Integrating (\ref{EulerSp}) with the ansatz (\ref{ansatzR}) with respect to $\rho$, results in 
%The radius $R(z)$ minimizes the functional
%
\begin{equation}
\int_{-t/2}^{t/2}\left[I_0\left(\frac{dR}{dZ}\right)^2+I_1R^2\left(\frac{\pi^2}{16\varkappa^2}\right)-I_2R-I_3hR^2\right]dZ
\label{EulerR}
\end{equation}
where
\begin{align}
&I_0=\int_0^1t^3\left(\frac{d\theta}{dt}\right)^2dt=\frac{\pi^2}{4}, I_1=\int_0^1t\sin^2\theta dt=\frac{1}{4},\nonumber\\
&I_2=\int_0^1t\left(\frac{d\theta}{dt}\right)dt=\frac{\pi}{2}, I_3=\int_0^1\left(\frac{t}{2}\right)\cos(\theta)dt=-\frac{1}{\pi^{2}}
\end{align}
The Euler equation for the functional (\ref{EulerR}) with the boundary conditions
\begin{equation}
\left(\frac{dR}{dZ}\right)_{Z=\pm t/2}=R(Z)\beta \frac{I_1}{I_0} 
 \end{equation}
has an analytical solution \cite{JETP00}:
\begin{equation}
R(Z)=\frac{1}{\pi A}\left(1-\frac{\beta\cosh(Z\sqrt{A})}{\beta\cosh(\frac{t\sqrt{A}}{2})+\pi^2\sqrt{A}\sinh(\frac{t\sqrt{A}}{2}}\right),\, A=\frac{1}{16\varkappa^2}+\frac{4 h}{\pi^4}. 
\end{equation}
%
 %$A=(4\varkappa)^{-2}+4\pi^{-4}h$ . 
%
Considered analytical approach for arrow-like angular profiles
describes numerically obtained solutions with
a good accuracy, and can be generally applied 
for other appropriately chosen initial functions $\theta(\rho/R(z))$.

\begin{figure}
\centering
\includegraphics[width=18cm,keepaspectratio]{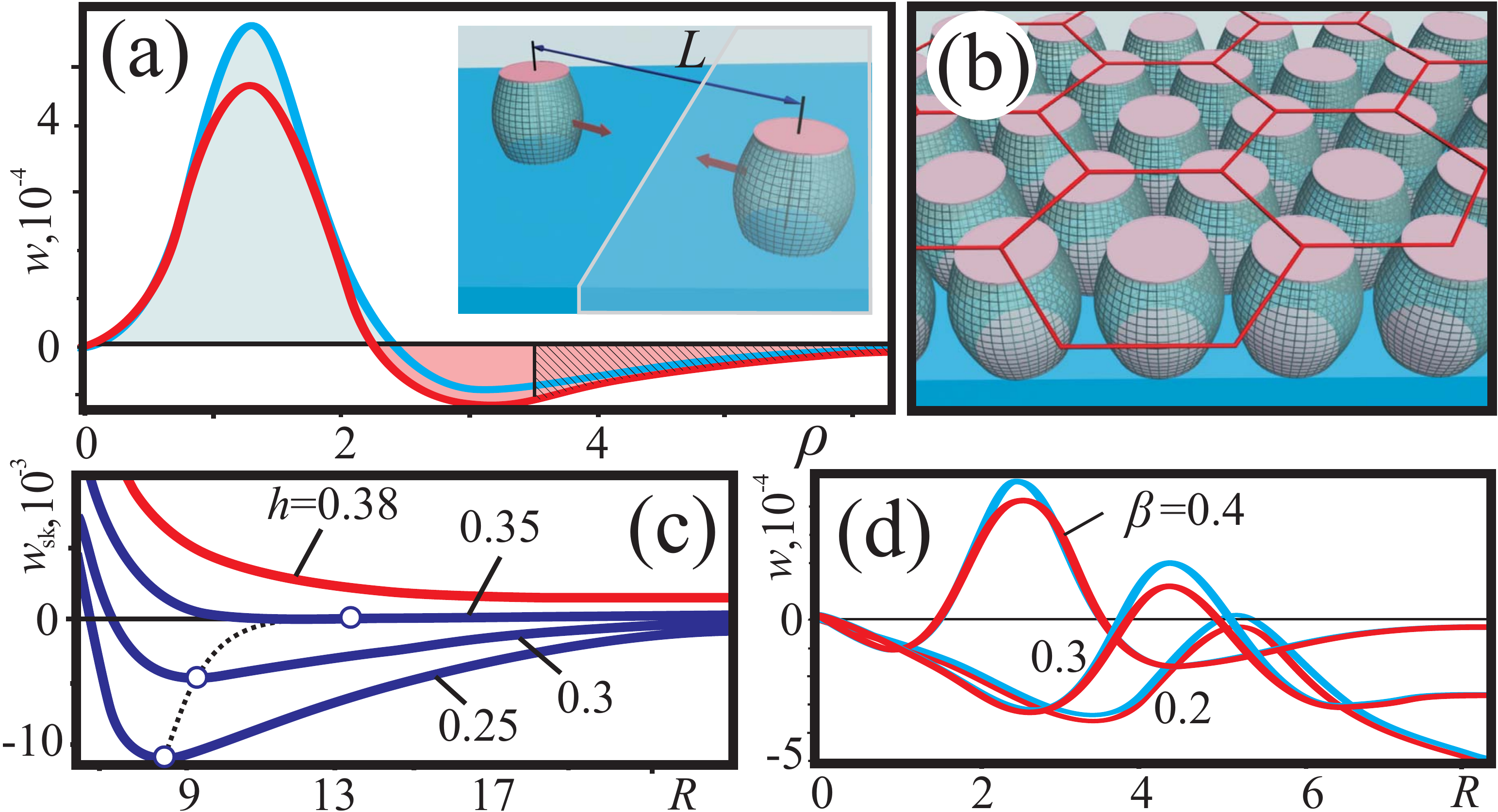}
\caption{
Distribution of energy density
in the isolated skyrmion state ($\beta=0.5$) (a)
and in the skyrmion lattice ($\beta=0.2$) (d) 
for the profiles $\theta(\rho)$
at the surface (blue line) and in the central film plane (red line)
($\varkappa=3, h=0$);
(b) schematic representation
of the hexagonal skyrmion lattice in the circular cell approximation; 
(c) dependence of the skyrmion
energy on the radius of lattice cell $R$ for different values
of the applied magnetic field ($\varkappa=3, \beta=0.1$). Above
the threshold field (red line) only isolated skyrmions can exist.
\label{attr}
}

\end{figure}

%%%%%%%%%%%%%%%%%%%%%%%%%%%%%%%%%%%%%%%%%%%%%%%%%%%%%%%%%%%%%%%%%%%%%%%%%%%%%%%%%%%%%%%%%%%%%%%%%%%%%%%%%%%%%%%%%%%%%%%%%%%
\section{Condensation of repulsive skyrmions into a lattice \label{CondensationSpherulites}}
%%%%%%%%%%%%%%%%%%%%%%%%%%%%%%%%%%%%%%%%%%%%%%%%%%%%%%%%%%%%%%%%%%%%%%%%%%%%%%%%%%%%%%%%%%%%%%%%%%%%%%%%%%%%%%%%%%%%%%%%%%%%%%

In the \textit{bulk} helimagnets the inter-skyrmion interaction is 
known to be repulsive and screened at large distances
$L$. 
In thin confined layers the standard interaction of skyrmions 
is modified, and related to the exponential decay of the
polar angle. 
The solution of the linearized Euler equation (\ref{EulerSp}) for $\rho\longrightarrow\infty$  has the asymptotic decay with
\begin{equation}
\theta({\rho},Z)\propto\cos{(\lambda Z)}\exp{(-\alpha\rho)}.
\end{equation}
Here, $\alpha=(\pi/4\varkappa)^2-h+\lambda^2$, and 
$\lambda$ are the roots of the transcendental equation,
$\lambda\tan{\lambda Z}=\beta$ (Fig. \ref{attr} (a), inset).
The energy density underlying such a slow rotation is negative and depends on the distance from the center of the layer. In Fig. \ref{attr} (a) I plotted the energy density distribution in the profile at the surface (blue line) and in the center of the layer (red line). 
The interaction energy of two vortices has the expression:
\begin{equation}
U(L,\beta)=\sqrt{\frac{2\pi L}{\alpha^3}}\exp{(-\alpha L)}.
\end{equation}
The parameter $\lambda$ due to the anchoring effect of the surfaces modifies this inter-skyrmion potential in comparison with the volume case. The lattice will be established from isolated skyrmions when the DM energy contribution (red-shaded area of energy distribution $w_{sk}(\rho)$, Fig. \ref{attr} (a)) of \textit{all} profiles $\theta(\rho,Z)$ outweights the exchange energy contribution (blue-shaded area). Since the exchange part of the energy density is much larger for the surface profiles $\theta(\rho,\pm t/2)$ (blue line in Fig. \ref{attr} (a)), the field of lattice formation will be smaller for larger values of parameter $\beta$ (see, for instance, the second plot in Fig. \ref{PD} (d), $\varkappa$ here is fixed). This explains the shape of the lability surface for the lattice of spherulites in Fig. \ref{PD} (a).
For Skyrmionic states in the lattice I used the circular-cell approximation \cite{Brandt03} and
solved Eq. (\ref{functional2}) with boundary conditions $\theta(0,Z)=\pi,\theta(R(Z))=0$ (Fig. \ref{attr} (b)). 
%
%Described mechanism of lattice formation through nucleation
%and condensation of isolated Skyrmions
%follows a classification introduced by DeGennes [11]
%for (continuous) transitions into incommensurate modulated
%phases.
%
%Such nucleation-type phase transitions are rather frequent
%in the condensed-matter physics and describe, for example, phenomena
%of vortex lines nucleation in type II superconductors,
%transitions between nematic and cholesteric liquid crystals,
%and/or origin of stripe/bubble domain structures from isolated 
%Bloch/Neel domain walls. 

%%%%%%%%%%%%%%%%%%%%%%%%%%%%%%%%%%%%%%%%%%%%%%%%%%%%%%%%%%%%%%%%%%%%%%%%%%%%%%%%%%%%%%%%%%%%%%%%%%%%%%%%%%%%%%%%%%%%%%%%%%%
\section{Skyrmion lattices versus helicoids. Phase diagram of solutions \label{spherulitesPD}}
%%%%%%%%%%%%%%%%%%%%%%%%%%%%%%%%%%%%%%%%%%%%%%%%%%%%%%%%%%%%%%%%%%%%%%%%%%%%%%%%%%%%%%%%%%%%%%%%%%%%%%%%%%%%%%%%%%%%%%%%%%%

\begin{figure}
\centering
\includegraphics[width=18cm,keepaspectratio]{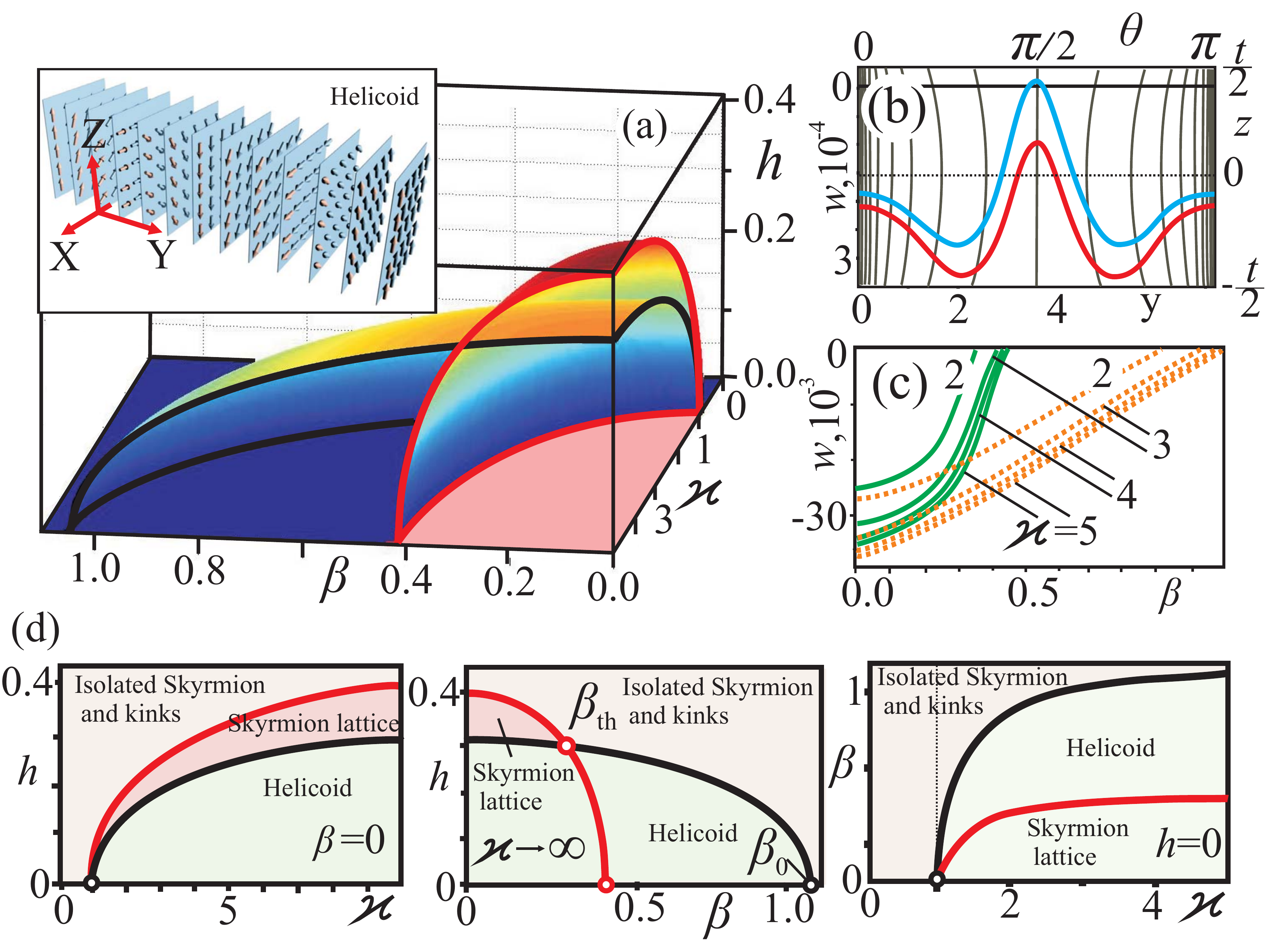}
\caption{(a) Phase diagram of solutions in the space
of control parameters $(\varkappa,\beta,h)$. Surfaces with
red and black lines bound the regions of skyrmion lattice
and helical state stability, respectively. (b) Distribution of the energy density in the
helicoid ($\beta = 0.5$) for the profiles $\theta(Y)$ (see inset in (a)
for schematic representation of distorted helicoid) at the surface (blue
line) and in the central film plane (red line) ($\varkappa = 3, h = 0$); thin black lines show the dependences $\theta(Z)$ in equidistant  planes $ZX$. 
(c) Energy dependence of equilibrium skyrmion lattice (green lines) and spiral state (dotted orange lines)
on the changed parameter of surface anchoring $\beta$ ($h=0$). (d) The two-dimensional cuts of the three-dimensional phase diagram (a).
\label{PD}
}

\end{figure}

Alternative to the two-dimensional skyrmion state
is the one-dimensional helicoid (Fig.\ref{PD} (a), inset) with
propagation vector along $Y$-axis, parallel to the surfaces. Angle $\theta$ of the magnetization  with respect to $Z$ axis in the helicoid can be obtained from the %The helicoid is characterized by the angle $\theta$ of the magnetization with respect to $Z$ axis which can be obtained 
% along $y$ axis, which
%is the solution of the 
Euler equation with boundary conditions:
\begin{align}
&\theta_{ZZ}+\theta_{YY}-\gamma\sin\,\theta\,\cos\,\theta =0,\nonumber\\
&\theta(0)=\pi, \theta(p)=0,\nonumber\\
&(\frac{\partial\theta}{\partial Z}+\beta\sin\theta\cos\theta)|_{Z=\pm t/2}=0
\end{align}
%
%with boundary conditions $\theta(0)=\pi, \theta(p)=0$, $(\partial\theta/\partial Z+\beta\sin\theta\cos\theta)|_{Z=\pm t/2}=0$.
%
where $p$ is a period of the helicoid. 
%$\theta$ is the angle of the magnetization with respect to $Z$ axis. 

In the distorted helicoid propagating in the magnetic layer, angle $\theta$ is a function of two coordinates, $Y$ and $Z$. This distinguishes the solutions $\theta(Y,Z)$ from those considered in chapter 4 with the dependence of the angle only on one spatial coordinate \cite{JMMM94}, $\theta=\theta(Y)$.
In the plane $XZ$ (see inset in Fig. \ref{PD} (a)) the solutions for helicoids have convex shape as it was noted also  for spherulites. I have plotted the dependences $\theta=\theta(Z)$ in equidistant planes with fixed coordinate $Y$ (thin black lines in Fig. \ref{PD} (b)). Such profiles have the most distorted shape in the vicinity of the point $Y=p/2$. Note that profiles for $Y=p/2$ are straight lines with  $\theta=\pi/2$: the magnetization is balanced by the two anchoring surface planes. Dependence of the angle $\theta$ along coordinate $Y$ is also strongly distorted by the anchoring effect of the surfaces: the order parameters
$\mathbf{m}$ (or $\mathbf{n}$) point up ($\theta=0$) and down ($\theta=\pi$) in the wide regions (domains) and rotate rapidly in the interstitial regions (domain walls). Such a behavior of the spiral state may be also deduced from the densest localization of lines $\theta(Z)$  in the vicinity of $\theta=0;\pi$ (thin black lines in Fig. \ref{PD} (b)).  % are not influenced by the 

%For $Y=p/2$ 
% for $Y=p/4$ and $Y=3p/4$, while for $Y=0,p/2,p$ the angle $\theta$ is almost constant in eaxh of the plane. 

%Rotational parts of the order parameter $\mathbf{m}$, $\mathbf{n}$ in the helicoid are inhomogeneously distorted and squeezed by the  anchoring in the plane $zy$. In the plane $xz$ the equilibrium solutions for helices also have a convex shape.
%
The distribution of the energy density in the helicoid along the propagation direction $Y$ is plotted in Fig. \ref{PD} (b) for the profiles at the surface $\theta(Y,t/2)$ (blue line) and in the center of the layer $\theta(Y,0)$ (red line). The largest loss of the rotational energy occurs for the planes with $\theta=\pi/2$ (Fig.\ref{PD} (b)). %as the magnetization is offset by the two anchoring surfaces (Fig.\ref{PD} (b)).

On the phase diagram the regions of modulated skyrmion matter and helicoidal state are bounded by the surfaces with red 
and black lability lines, respectively (Fig. \ref{PD} (a)). Corresponding two-dimensional cuts of this three-dimensional phase diagram are shown in  Fig. \ref{PD} (d).
For $h=0$ (i.e. in the plane $(\varkappa,\beta)$) the energy density of the helicoid is always lower than that for skyrmion lattice (Fig. \ref{PD} (c)), i.e. skyrmions are  metastable states with respect to helicoids for all possible values of the anchoring parameter $\beta$. The region of the thermodynamical stability of the helical state is marked by blue color in Fig. \ref{PD} (d). According to the phase diagram of Fig. \ref{PD} (d), increasing parameter $\beta$ (for fixed value of $\varkappa$) leads to the suppression of skyrmions (red line) for much lower values than the suppression of helical states (black line). If the skyrmion lattice as metastable state has been formed for $\beta=0$, then on the lability (black) line it releases free isolated skyrmions. But since the helicoids are the global minimum of the system above the lability line of the skyrmion lattice, the isolated skyrmions must undergo ellongation into the helices.

For $\varkappa\longrightarrow\infty$ (meaning zero uniaxial anisotropy \cite{JMMM99}), i.e. on the plane $(\beta,h)$ (Fig. \ref{PD} (d)), the thermodynamical stability of skyrmions can be achieved in the applied magnetic field only for relatively  small values of the anchoring parameter $\beta$ - in the triangular region marked by red color in Fig. \ref{PD} (d). For values of $\beta$ larger than some threshold $\beta_{th}$ (Fig. \ref{PD} (d)), the helicoids are the thermodynamically  stable state. 
On the basis of competing Zeeman energy and surface-induced uniaxial anisotropy the anchoring can be classified as weak (favouring skyrmion lattice in the field for $\beta<\beta_{th}$), intermediate (favouring helical state in zero and the applied magnetic field for $\beta_{th}<\beta<\beta_{0}$), and strong (only isolated domain walls and localized skyrmions can exist, $\beta>\beta_0$).

 %The anchoring  the anchoring effectively suppresses Skyrmion lattices, and localized Skyrmion solutions undergo elliptical instability signalling transformation into helical states (in the region between lability surfaces of helicoid and Skyrmion lattice).

%For small values of magnetic field $h$ the energy density of the helicoid is always lower than that for Skyrmion lattice (Fig. \ref{PD} (c)):  the anchoring effectively suppresses Skyrmion lattices, and localized Skyrmion solutions undergo elliptical  nstability signalling transformation into helical states (in the region between lability surfaces of helicoid and Skyrmion lattice).
%
%Applied magnetic field $h$ for small anchoring parameters, on the contrary stabilizes the Skyrmion lattice and  destroys helicoidal modulations.
%

%\textit{Some general words.}

%
%For liquid crystals other surface interactions (for example, $K_{13}, K_{24}$) can be included in the boundary contitions which may give additional impact on realised modulated states.

\section{Magnetostatic problem for isolated skyrmions \label{SkMagnetostatic}}

In previous sections I discussed characteristic features of modulated states arising as a result of competing internal
short-range interactions (as the exchange interactions, the Dzyaloshinkii-Moriya coupling, different types of intrinsic and induced magnetic anisotropy) and ignored the effects imposed by magnetodipole forces. In many magnetic nanostructures this assumption is justified by weakness of stray field effects as compared to the internal magnetic interactions (e.g.
nanosystems with in-plane magnetization). However, in a large group of nanostructures with perpendicular anisotropy the magnetodipole coupling play an important role to stabilize specific multidomain patterns and topological defects \cite{Kiselev07,Kiselev08,Hellwig07,Bran09,Kiselev08b}. Generally in nanolayers and multilayers with perpendicular magnetization the stray fields can also strongly modify chiral patterns - helices and skyrmions \cite{Kiselev11}.
In this section I investigate the  influence of long-range magnetostatic interactions on the characteristic features of isolated skyrmion states. It is only the first step to address the problem of the formation of the equilibrium modulated patterns under the competing influence of the Dzyaloshinskii-Moriya and magnetostatic interactions. 
%
%Theoretical analysis of chiral modulations requires the extension of the  model by including into the phenomelogical energy

\begin{figure}
\centering
\includegraphics[width=18cm,keepaspectratio]{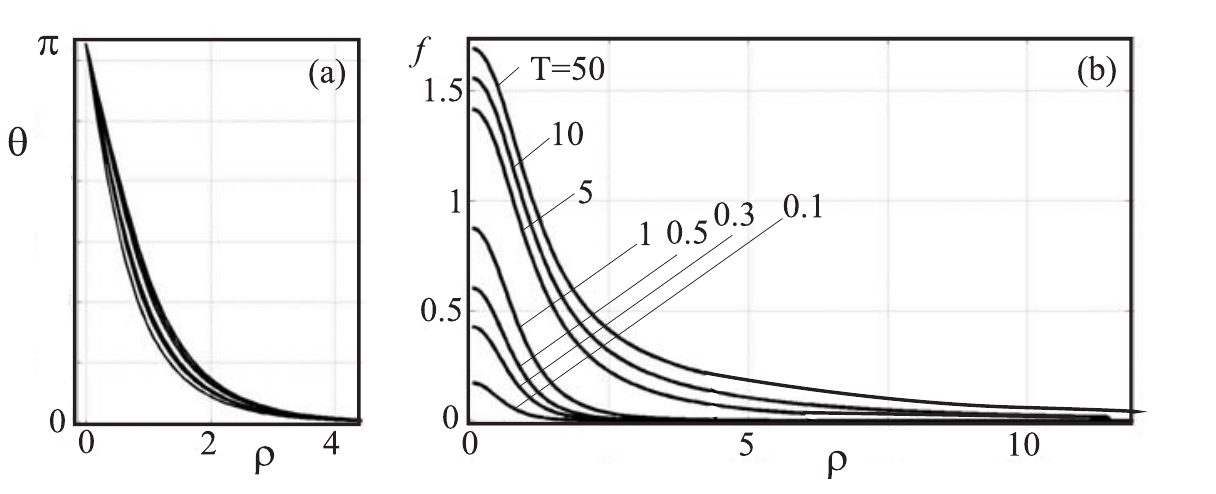}
\caption{(a) Equilibrium angular profiles $\theta(\rho)$ for skyrmion solutions of Eq. (\ref{demFunctional}) plotted for different values of film thickness $T$, $h=-0.05, \varkappa=0.5, Q=5$; (b) function $f(\rho)$ (\ref{frho}) for corresponding skyrmion solutions. 
}
\label{SkDem}
\end{figure}

The total energy $W$ of the skyrmion filement in the layer of thickness $T$ can be written in the following reduced form:
\begin{align}
\frac{W L_B}{2K_u}=&2\pi T \int_0^{\infty}\left[\left(\frac{d\theta}{d\rho}\right)^2+\frac{\sin^2\theta}{\rho^2}+\sin^2\theta+(2h-\frac{1}{Q})(1-\cos\theta)+\right.\nonumber\\
&\left.+\frac{4\varkappa}{\pi}\left(\frac{d\theta}{d\rho}+\frac{\sin\theta\cos\theta}{\rho}\right)-\frac{f(\rho)}{QT}\cos\theta\right]\rho d\rho 
\label{demFunctional}
\end{align}
where $L_B=\sqrt{A/K_u}$ serves as a scale of the non-dimensional radial variable $\rho$, and $Q=K_u/2\pi M^2$ is a quality factor, $\varkappa=\pi D/4\sqrt{AK_u}$. 

The stray-field energy of the skyrmion string had been derived by solving the corresponding magnetostatic problem as shown in Ref. \cite{Lin71,Tu71} for magnetic bubble domains.  Here, in Eq. (\ref{demFunctional}) 
\begin{equation}
f(\rho)=\int_0^{\infty}\int_0^{\infty}(1-\cos\theta(\rho'))(1-e^{-\xi T})J_0[\xi\rho']J_0[\xi\rho]\rho'd\rho'd\xi
\label{frho}
\end{equation}
$J_0$ is a Bessel function of zero order. 

I solve the magnetostatic problem for two types of localized solutions (Fig. 4.14 (a), (e)). In the absence of demagnetizing fields the first localized solution (Fig. 4.14 (a)) represents skyrmion stabilized by the DM interactions. The second type of localized solution (Fig. 4.14 (e)) is unstable and can exist only in the field applied along the magnetization in the center. In the presence of dipole-dipole interactions this second type of the localized solution becomes stable, it is the solution for the famous magnetic bubble domain \cite{Kiselev11}. 

Typical solutions $\theta(\rho)$ for localized skyrmions are shown in Fig. \ref{SkDem}. For quality factor $Q=5$ and $\varkappa=0.5$ I plot the set of solutions parametrized by the film thickness $T$. With the decrease of the thickness the angular profiles become more localized. The solutions for skyrmions can exist in very strong positive field: they are protected from the collapse by DM interaction. 

\begin{figure}
\centering
\includegraphics[width=18cm,keepaspectratio]{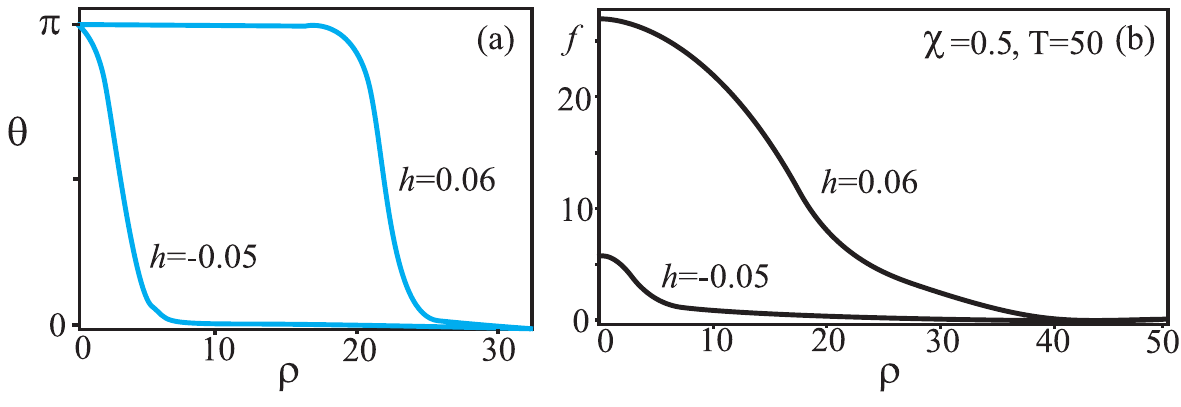}
\caption{ (a) Equilibrium angular profiles $\theta(\rho)$ for magnetic bubble domains (Eq. (\ref{demFunctional})) plotted for $h=-0.05$ and $h=0.06$, and $\varkappa=0.5, Q=5$; (b) function $f(\rho)$ (\ref{frho}). 
}
\label{BubbleDem}
\end{figure}

Solutions for magnetic bubble domains shown in Fig. \ref{BubbleDem}  are characterized by much larger radial sizes in comparison with skyrmions. Such solutions are stable only in the narrow interval of magnetic field applied against the magnetization in the center. Any slight deviation of the control parameters $T$, $\varkappa$, $Q$, and $h$ leads to the instability of these magnetic domains, although as a solution of the Euler equation bubble domains can exist even in the absence of dipole-dipole interaction (as it is seen from Fig. 4.14 (e)-(h)).

Bubble domains and skyrmions are two fundamentally different types of solutions of micromagnetic functional. As bubble domains are stabilized only in the narrow region of applied magnetic fields and film thicknesses \cite{Hubert98}, the skyrmions look more preferable for the possible applications. 

%%%%%%%%%%%%%%%%%%%%%%%%%%%%%%%%%%%%%%%%%%%%%%%%%%%%%%%%%%%%%%%%%%%%%%%%%%%%%%%%%%%%%%%%%%%%%%%%%%%%%%%%%%%%%%%%%%%%%%%%%%%%
%old
%\section{Observations of  skyrmionic and helical textures in Fe$_{0.5}$ Co$_{0.5}$Si nanolayers: theoretical analysis \label{ObservationAnalysis}}

%new
\section[Observations of  skyrmionic and helical textures in Fe$_{0.5}$ Co$_{0.5}$Si nanolayers]{Observions of  skyrmionic and helical textures in Fe$_{0.5}$ Co$_{0.5}$Si nanolayers: theoretical analysis \label{ObservationAnalysis}}
%%%%%%%%%%%%%%%%%%%%%%%%%%%%%%%%%%%%%%%%%%%%%%%%%%%%%%%%%%%%%%%%%%%%%%%%%%%%%%%%%%%%%%%%%%%%%%%%%%%%%%%%%%%%%%%%%%%%%%%%%%%%%%

\begin{figure}
\centering
\includegraphics[width=18cm,keepaspectratio]{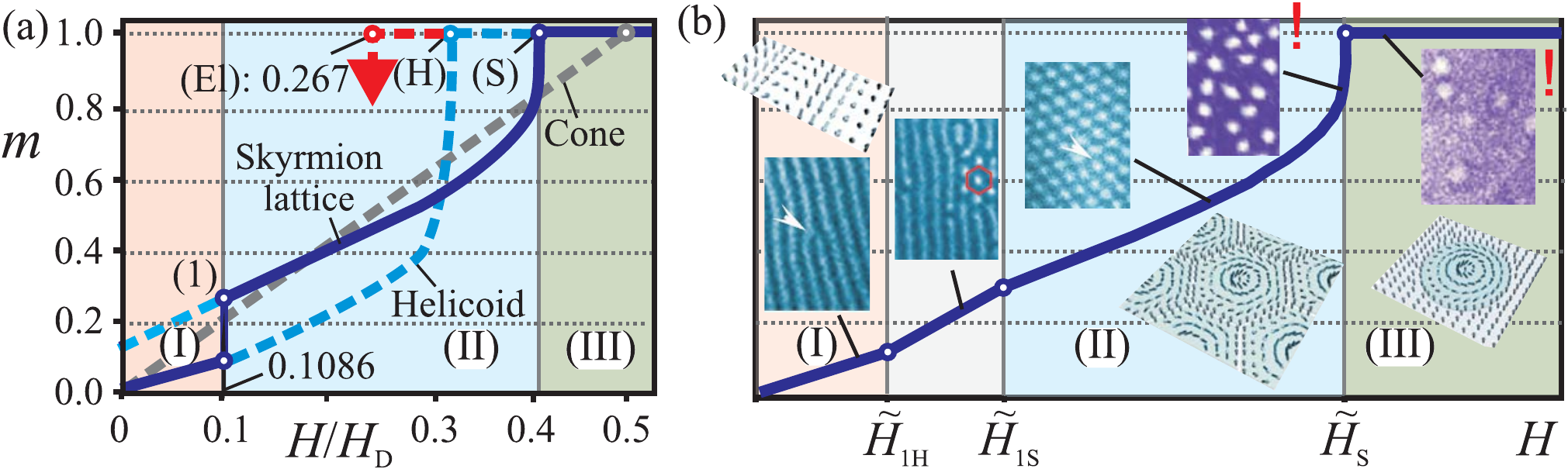}
\caption{ The ideal magnetization curves
for a bulk sample (based on results of
\cite{JMMM94}) (a) and for a thin layer (b)
of a cubic helimagnet with
suppressed cone phases.
Solid lines indicate the thermodynamically
stable states; dashed lines in Fig. (a), metastable
configurations.
The transition field $H_1$ in the thin layer
"spreads" into a region of multidomain
states.
Fragments of experimentally observed images
\cite{Yu10} demonstrate excellent agreement
with theoretically calculated magnetization 
curves. The patterns indicated with sign (!)
contain images of isolated chiral skyrmions.
}
\label{mcurves}
\end{figure}

\vspace{1mm}
Real-space images of skyrmion states 
in a thin layer of cubic helimagnet Fe$_{0.5}$Co$_{0.5}$Si
have recently been obtained by using Lorentz transmission
electron microscopy \cite{Yu10}. 
This is the first clear experimental
manifestation of \textit{chiral skyrmion states} in a non-centrosymmetric magnetic crystal.
The first-order transition
of a helicoid into a skyrmion lattice and  its subsequent transformation
into a system of isolated skyrmions  observed in bias magnetic
fields (Figs. 1, d-f, 2, 3 (a-d) in \cite{Yu10})
are in excellent agreement with the theoretical
predictions on the behavior of skyrmions and the field-driven 
transitions into densely packed skyrmion lattices
according to the magnetic phase diagrams calculated
earlier \cite{JMMM94,Nature06} (Fig.~\ref{mcurves}).

In the experiments, the thickness 20~nm of 
the magnetic layer Fe$_{0.5}$Co$_{0.5}$Si 
is much smaller than the helix period $L_D$ = 90 nm \cite{Yu10}.
But, even for such a small thickness, the conical state  
propagating only for a fraction of a period perpendicularly
through the layer has lower energy than a skyrmion lattice, 
absent additional effects that stabilize it in applied fields. 
Usually in magnetic nanolayers strong perpendicular 
uniaxial anisotropy arises, either as a result of surface 
effects \cite{Johnson96} or of lattice strains.
Thus, induced anisotropies give a possible explanation 
for the experimental observation of the skyrmions in
these Fe$_{0.5}$ Co$_{0.5}$Si layers, in accordance with
the phase diagram for cubic helimagnets with uniaxial 
distortions [XI].

Fig.~\ref{mcurves} presents the magnetization curve for a bulk isotropic
helimagnet (a) (based on results of \cite{JMMM94}, Fig. 12)
and the corresponding magnetization curve for a thin layer
involving demagnetization effects \cite{Hubert98} (b). 
Compared to theoretically calculated
values in a bulk material ($H_S$, $H_H$) the
corresponding critical fields in a thin layer
are shifted, and their values can be estimated
as $\widetilde{H}_{S(H)} = H_{S(H)} + 4\pi M$.
Due to demagnetization effects multidomain states
can be stabilized in the vicinity of the transition 
field $H_1$ \cite{JMMM94}. The boundaries of these 
regions with coexisting phases can be estimated as
$\widetilde{H}_{1H}= H_1 + 4\pi M m_{\small{H}}(H_1) $,
$\widetilde{H}_{1S}= H_1 + 4\pi M m_{\small{H}}(H_1) $.
The magnetizations of the competing phases at the transition field
equal  $m_{\small{H}}  (H_1)$ = 0.111 and $m_{\small{S}}  (H_1)$ = 0.278.
The jump of the magnetization at the transition equals
$\Delta M = [ m_{\small{S}} (H_1) - m_{\small{H}} (H_1)] M = 0.167 M$,
i.e., it reaches about 17 \% of the saturation value.  

The magnetization curves in Fig.~\ref{mcurves} are constructed
for ideally soft magnetic material under the condition that only
the equilibrium states are realized in the magnetic sample.
In real materials the formation of the equilibrium 
states  is often hindered (especially during the phase transitions), 
and evolution of metastable states and hysteresis effects play 
an important role in the magnetization processes.
Particularly, the formation of the skyrmion lattice below $H_S$
can be suppressed. Then isolated skyrmions exist below this
critical field. At a critical field H$_{El}$
the skyrmions become unstable with respect to elliptical 
deformations and "strip-out" into isolated 2$\pi$ domain walls.
In a bulk material $H_{El} = 0.267 \, H_D$ 
(indicated in Fig.~\ref{mcurves} with a red arrow). 
In a thin layer, one estimates
$\widetilde{H}_{El} = H_{El} + 4 \pi M$.
As discussed earlier \cite{JETP89,Nature06,JMMM94}
the evolution of chiral skyrmions in magnetic fields 
has many features in common with that of bubble domains
in perpendicular magnetized films,\cite{Hubert98} and
with Abrikosov vortices in superconductors \cite{Blatter94}.

The images from Ref. \cite{Yu10} (Fig.\ref{mcurves} (b))
reflect in details theoretically
predicted evolution of the chiral modulations
in the applied magnetic field:
the helicoid phase is realized at low fields (region (I));
at higher field this transforms into the skyrmion lattice
(region (II)) via an intermediate state
($\widetilde{H}_{1H}< H < \widetilde{H}_{1S}$);
finally the skyrmion lattice by extension of the
period transforms into the homogeneous phase 
where isolated skyrmions still exist as topologically
stable 2D solitons.

Two patterns indicated in Fig.~\ref{mcurves} (b)
with exclamation mark manifest the main result
of Ref.~\cite{Yu10}:  the first images of 
\textit{static two-dimensional localized states}
aka \textit{chiral skyrmions}!
In Ref.  \cite{Yu10} this result has been overlooked
and misinterpreted as a coexisting ferromagnetic 
and skyrmion lattice phases.
As it was expounded in the previous section, the
transition of the skyrmion lattice into the 
homogeneous state is a \textit{continuous} transition,
but of the particular \textit{nucleation} type.
Such transitions exclude the formation of coexisting states.

The condensed skyrmion phases in the micrograph of  Ref.~\cite{Yu10}
also appear as heavily distorted densely packed two-dimensional
lattice configurations. This is expected for skyrmionic matter. 
As these mesophases are composed from elastically
coupled radial strings, dense skyrmion configurations 
generally do not form ideal crystalline lattices 
but various kinds of partially ordered states, 
e.g. hexatic ordering implying only orientational order
of bonds without positional long-range order, or other 
glassy  arrangement following standard arguments 
put forth for the similar vortex matter in type-II superconductors
\cite{Blatter94}.
The observation derives from the particle-like (or string-like)
nature of skyrmions and suggests that skyrmionic mesophases may
display rich phase diagrams.

\section{Observation of skyrmion states in chiral liquid crystals}

\begin{figure}
\centering
\includegraphics[width=18cm,keepaspectratio]{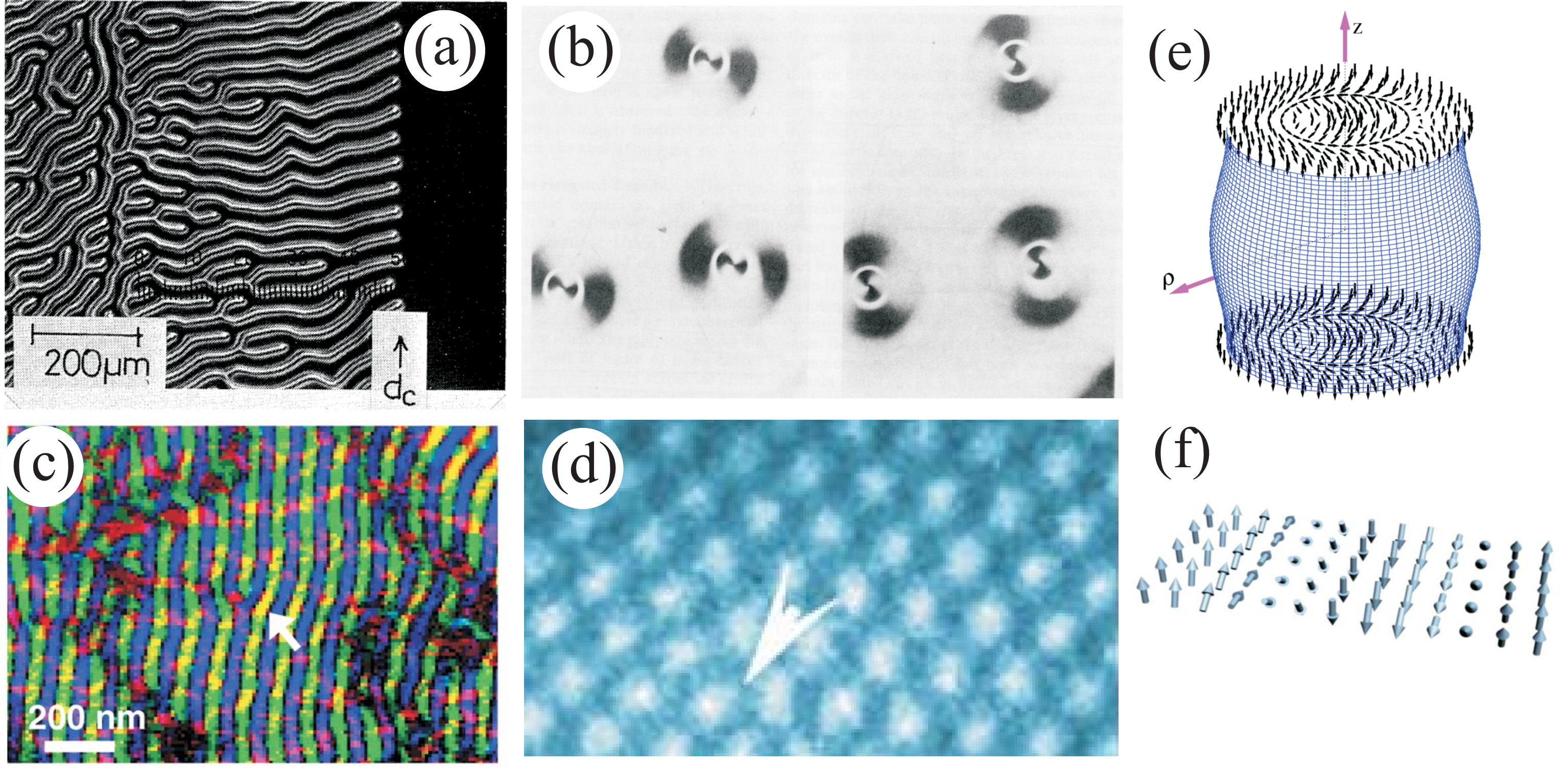}
\caption{Modulated phases in liquid crystals (a),(b) and their analogues in magnetic systems (c),(d):
(a), (c) helical modulations in  cholesterics \cite{Oswald00} and in cubic helimagnets \cite{Uchida08} with the distribution of the order parameters shown as a sketch in (f); (b), (d) isolated spherulites in a chiral liquid crystal \cite{Stieb80} and hexagonal lattice of chiral skyrmions in (Fe,Co)Si \cite{Yu10}. (e) shows schematically the structure of a spherulite. %; (c) chiral helices in a cubic helimagnet \cite{Uchida08}; (d) hexagonal lattice of chiral skyrmions in (Fe,Co)Si \cite{Yu10}.
\label{LCobservation}
}
\end{figure}

As it was noted in section \ref{spherulitesPhen}, the phenomenological energy of cholesterics (\ref{lc1}) has an identical mathematical form as that of cubic helimagnets. %Therefore, chiral textures must be considered in the close relation to each other in both condensed matter systems. 
This implies close relations between chiral textures in both condensed matter systems. However, in contrast to magnetic systems favoring smooth distributions of the order parameter, liquid crystals usually form patterns composed of various types of singularities. Defects in liquid crystals are of various dimensionalities, not only line defects, but also points and walls, and appear due to the prevalence of orientational order over positional in the applied magnetic or electric fields. In the defects the director $\mathbf{n}$ is said to be undefined and the properties of defects are often not well controlled. 
For many years the investigations of liquid crystal textures have been mostly concentrated on the processes related to the formation and evolution of these
topological defects \cite{Wright89,Oswald00,Hornreich82}. Only during the last decade physical analogies between liquid
crystal and magnetic systems have been utilized to find new skyrmionic textures in these systems \cite{Nature06,JETP00,Bogdanov03a,Bogdanov98c}. Particularly, the analogy with cholesteric blue phases \cite{Hornreich82} has been used to establish the skyrmionic ground state in chiral magnetic metals near the ordering temperature \cite{Nature06}. The results on observations of specific skyrmion states (spherulitics) in confined cholesteric systems (Fig. \ref{LCobservation} (b) \cite{Oswald00}) can help to investigate similar structures in magnetic nanolayers \cite{Yu10,Yu10a}. Liquid crystals have several advantages over magnetic systems for the modelling and investigation of various inhomogeneous structures. The system parameters can be varied over wide limits to establish necessary conditions for a given experiment; as a rule experiments are conducted at room temperature and are comparatively simple; the results of investigations are easily visualized, to a degree not usually attainable in the investigation of magnetic nanolayers. 

Fig. \ref{LCobservation} shows modulated patterns in chiral magnets (c,d) in comparison with those in chiral liquid crystals (a,b). Under non-restricted conditions, chiral-nematic LC molecules, which are characterized by an antisymmetric center in the molecule, organize themselves by following a helicoidal director alignment (Fig. \ref{LCobservation} (a)). However,  more exotic ground states are possible (for example, cholesteric fingers \cite{Smalyukh05} and/or triple-twist torons \cite{Smalyukh09}), in particular, spherulites (Fig. \ref{LCobservation} (b)). %Among them are cholesteric fingers \cite{Smalyukh05}, triple-twist torons \cite{Smalyukh09}, and/or spherulites. %For example, different types (one distinguishes four types of fingers) cholesteric fingers can nucleate with very different static and dynamical properties during the nematic-cholesteric phase transition with variable ratio $C=t/p$. Only the fingers of fir

Such circular objects, called bubble domains or spherulites, can be formed as isolated entities or arranged in a hexagonal array (Fig. \ref{LCobservation} (b)). For the first time, they  have been observed  in 1974 simultaneously by Kawachi and Kogure \cite{Kawachi74} and Haas and Adams \cite{Haas} in materials of negative dielectric anisotropy. The spherulites were generated by applying pulses of DC or AC low frequency electric field strong enough to induce electrodynamic turbulence. Two years later, Bhide et al. \cite{Bhide76} studied the optical properties of this pattern by laser diffraction and proposed that the bubble domain was a cholesteric pocket with oblate spheroid shape. This model was replaced by a more convincing model in papers of Akahane et al. \cite{Akahane77,Hirata81}. In this model which was inspired by the paper of Cladis and Kleman \cite{Cladis72}, two looped disclinations were assumed to exist near the glass plates. Another model of Stieb \cite{Stieb80} suggested that only one singular line is located along the axis of spherulite (Fig. \ref{LCobservation} (b)). As well, according to the experiments of Pirkl et al. \cite{Pirkl93} the bubble domains can be formed from the looped finger in the cholesterics with positive dielectric anisotropy. In the present chapter I have shown that a non-singular model with a continuous distribution of the director field in the spherulite is among the solutions of the equations minimizing the Frank functional for the cholesteric layer with the homeotropic boundary alignment. Thus, the spherulitic bubbles in anchored chiral liquid crystal films \cite{Kawachi74,Haas,Bhide76} may be skyrmion textures. 

%Cladis and Kleman \cite{}.

\section{Conclusions}

In this chapter I investigated some effects imposed by the confined geometry of magnetic nanolayers on skyrmion states, namely, the influence of surface-induced anisotropy and demagnetizing fields on the stability and the structure of localized skyrmions. I showed that the surface-induced anisotropy produces pinning (anchoring) effect on magnetic states. It suppresses very effectively modulated skyrmion states and distorts the tubular structure of skyrmion filements making them of convex-like shape with narrow "necks" near the surface. The similarity of phenomenological models for nonsingular spherulitics in liquid crystals and skyrmions in chiral magnetic materials  offers new prospects for investigations of these solitonic states within a common theoretical approach. The rich experimental material on observation of spherulitic patterns in liquid crystals \cite{Oswald00} in this sense can be used as guidelines for investigations of skyrmion states in chiral magnetic layer systems.

Also in this section I addressed the problem of the influence of demagnetizing effects on the skyrmion solutions. I showed that skyrmions and magnetic bubble domains are two different branches of cylindrical magnetization structure with different mechanism of internal stabilization.

%\textbf{With the new work [Yu10], that is a real experimental breakthrough in the field, our theoretical work has acquired additional importance and it is still timely. This new experimental work is unequivocal proof for the existence of Skyrmionic states, including isolated free Skyrmions, at low temperatures: the subject of this manuscript. Induced anisotropies and/or lattice distortions are likely a major contributing factor to stabilize these states, as seen in these thin single-crystalline plates.}

\end{document}